\documentclass[aps,prl,twocolumn,a4paper,superscriptaddress]{revtex4-1}

\usepackage{graphicx,color,amsfonts,amsmath,amssymb,bm,mathrsfs,latexsym,setspace}

\newcommand\rme{{\rm e}}

\newcommand{\eg}{{\it e.g.}}
\newcommand{\ie}{{\it i.e.}}
\newcommand{\etal}{{\it et al.}}

\newcommand{\ld}{\ell_{\rm D}}
\newcommand{\bld}{\ell_{\sigma}}

\newcommand{\eps}{\varepsilon}
\newcommand{\kb}{k_{\rm B}}

\newcommand{\MP}{{\rm MP}}
\newcommand{\HW}{{\rm HW}} 

\begin{document}

\noindent {\bf Colloidal  
aggregation and critical Casimir forces}

A recent Letter~\cite{Bonn} reports the experimental observation of 
aggregation of colloidal particles dispersed in a
liquid mixture of heavy water (\HW) and 3-methylpyridine (\MP). The authors claim that the experimental data can be explained in terms of a 
model which accounts solely for the competing effects of the inter-particle electrostatic repulsion and the attractive critical Casimir force, the decay lengths of which are set by the Debye screening length $\ld$ and the correlation length $\xi$ of the liquid mixture, respectively. 
Here we show, however, that the reported aggregation actually occurs 
within ranges of values of  $\xi$  and  $\ld$  ruled out by the proposed model and that most of the experimental data presented in Fig.~4(c) of Ref.~\cite{Bonn} cannot be consistently interpreted in terms of such a model.

Following Ref.~\cite{Bonn}, the effective 
potential $V(l)$ of the force acting on two identical 
colloids of radius $R=0.2\,\mu$m and surface charge $\sigma=2.72\,\mu\mbox{C}/\mbox{cm}^2$ consists of an electrostatic part $V_{\rm el}(l)= [2\pi R\sigma^2 \ld^2/(\eps\eps_0)] \rme^{-l/\ld}$ plus the critical Casimir part 
$V_{\rm C}(l)= -\kb T \pi A\, (R/\xi)\rme^{-l/\xi}$~\cite{long}, where $l$ is the 
surface-to-surface distance
of the two colloids, $\eps$ the dielectric constant of the mixture, $A=1.2$--$1.5$ a universal constant~\cite{long} 
(set to $2$ in Ref.~\cite{Bonn}), 
and $T$ the temperature. 
The expressions for $V_{\rm el}$ and $V_{\rm C}$ (valid within the Derjaguin approximation $l\ll R$~\cite{isr,long}) have been derived assuming $l \gtrsim \ld$ 
and $l \gtrsim \xi$, respectively, 
with $\xi \simeq\xi_0(1-T/T_c)^{-\nu}$ and $\nu\simeq 0.63$, which is expected to hold for $\xi \gg \xi_0$, \ie, sufficiently close to the critical point at temperature $T_c$.
For the present mixture $T_c\simeq 310\,$K, $\xi_0\simeq 3\,$\AA~\cite{mu}, and therefore significant (non-universal) corrections to $V_C$ are expected to occur within the gray area in the figure,
corresponding to $\xi \lesssim 5\,\xi_0$.
Adding NaCl to the mixture causes both a reduction of $\ld$~\cite{Bonn} and a screening of 
$\sigma$.   
The effective value $\sigma^*$ which replaces $\sigma$ in $V_{\rm el}$ can be determined via the Grahme equation~\cite{isr}, which yields $\sigma^* = \sigma\, g(\ld/\bld)$ where $g(\rho)=2/(1 + \sqrt{1 + \rho^2/4})$ and $\bld\equiv \kb T\eps\eps_0/(e\sigma)\simeq 0.2\,$nm for a near-critical temperature 
$T\simeq T_c $  
and $\eps \simeq 26$. 
This value of $\eps$ for the 
mixture with critical \MP\ mass fraction  $x_c\simeq 0.29$~\cite{mu} 
(\ie, volume fraction $\phi_c\simeq 0.32$~\cite{densities}) is estimated 
on the basis of the dielectric constants of pure \HW\ (78.3) and \MP\ (10.0) at 
$T\simeq298\,$K~\cite{lb} via the Clausius-Mossotti formula, neglecting 
the fractional volume change (see, \eg, Ref.~\cite{lpb}). 
The various forms  of the resulting potential $V$ are sketched in the figure for each of the 
six regions delimited by the thin solid lines in the $\xi$--$\ld$ plane. 
\begin{figure}
\begin{center}
\includegraphics[scale=0.72]{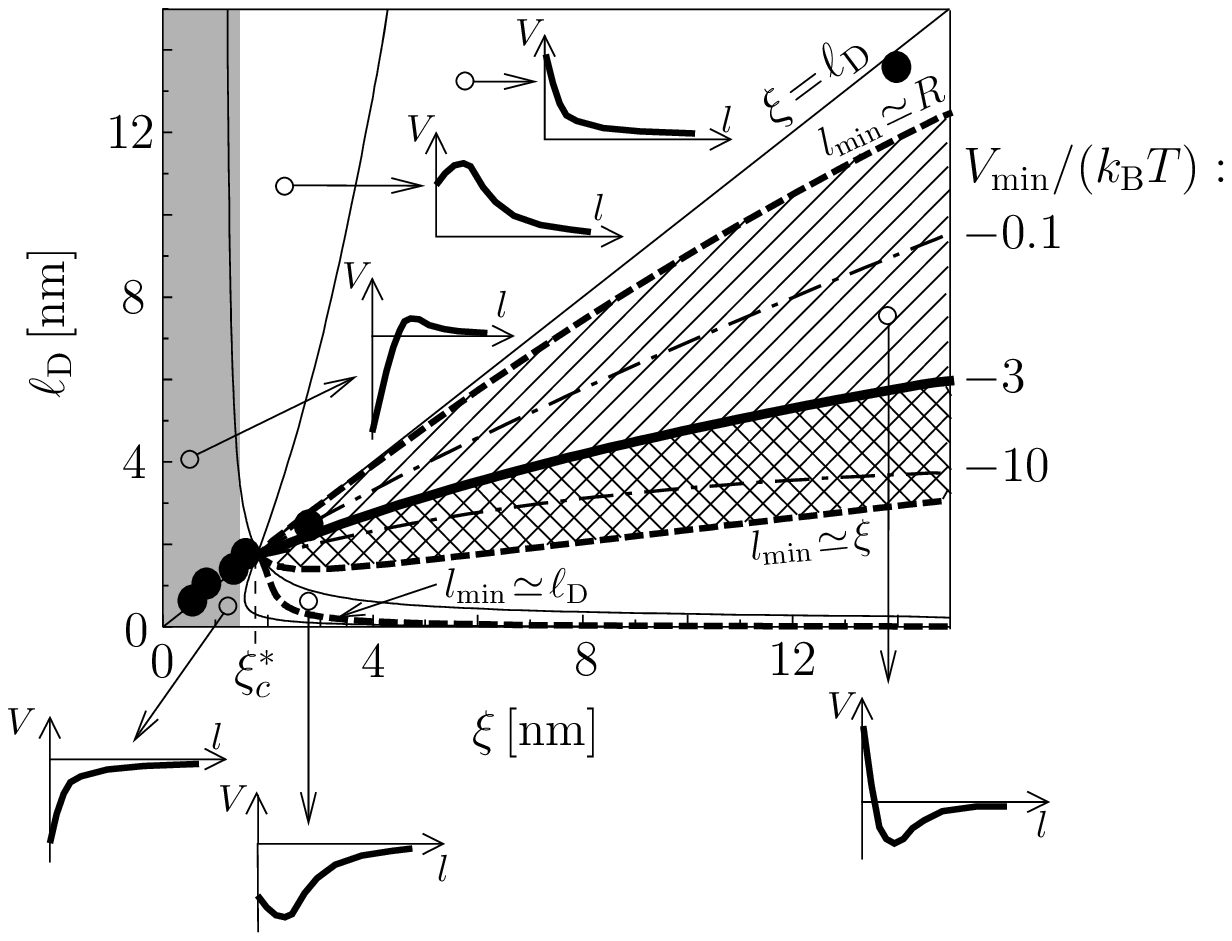}\vspace{-8mm}
\end{center}
\label{fig:model}
\end{figure}
All these lines meet for 
$\ld=\xi=\xi_c^*$, where we introduce  
$\xi_c \equiv [Ae\bld/(2\sigma)]^{1/3} \simeq 1\,$nm and define 
$\xi_c^*$ via $(\xi^*_c/\xi_c)^3 g^2(\xi^*_c/\bld)=1$ which renders $\xi^*_c\simeq 1.8\,$nm for the present experiment. 
In order for the above model of $V=V_{\rm el} + V_{\rm C}$ to be consistent, it is necessary that the distances typically sampled by the aggregating colloids, i.e., close to  the position  $l_{\rm min}$ of the minimum of $V(l)$, are in the range $R \gg l_{\rm min}\gtrsim \xi,\,\ld$ so that the above expressions for $V_{\rm el,C}$ are applicable. These three conditions, delimited by the thick dashed lines in the figure,
are satisfied within the enclosed hatched area. In addition, the potential depth $V_{\rm min}\equiv V(l_{\rm min})$ --- which, for fixed $\xi$, gradually increases upon decreasing $\ld < \xi$ [see the solid and dash-dotted lines within the hatched area, corresponding from top to bottom to $-V_{\rm min}/(\kb T) = 0.1,3,10$] has to be large enough to cause the aggregation of the particles. This requires $V_{\rm min}\lesssim - E_{ACT} \simeq -3\kb T$~\cite{Bonn}. 
Within the hatched area, this condition is fulfilled only in its cross-hatched part, which
is significantly smaller than the region $\xi<\ld$ indicated in Ref.~\cite{Bonn} and therefore it does no longer agree with the experimentally determined aggregation line~\cite{Bonn} (data points in the figure).  

Thus a careful study of the model proposed in Ref.~\cite{Bonn} shows that it does not predict aggregation to occur at 
$\xi=\ld$, leading to a discrepancy with the reported data. Such a discrepancy is particularly 
significant at values of $\xi\gg\xi_0$ for which an interpretation in terms of universal critical phenomena could be acceptable.

%
%
\vspace{2mm}
\noindent 
Andrea Gambassi${}^1$ and S.~Dietrich${}^2$\\
\noindent 
\tighten{\small${}^1$SISSA -- International School for Advanced Studies and INFN, I-34151 Trieste, Italy\\
${}^2$Max-Planck-Institut f\"ur Metallforschung and Institut f\"ur Theoretische und Angewandte Physik der Universit\"at Stuttgart, D-70569 Stuttgart, Germany}

\end{document}